# Quantifying the relative importance of the spatial and temporal resolution in energy systems optimisation model

Nandi Moksnes,[1,*] and William Usher[1]
[1] Department of Energy Technology, KTH Royal Institute of Technology Stockholm, Brinellv. 68, Stockholm, Sweden

*Correspondence: nandi@moksnes.se

## SUMMARY

An increasing number of studies using energy system optimisation models are conducted with higher spatial and temporal resolution. This comes with a computational cost which places a limit on the size, complexity, and detail of the model. In this paper, we explore the relative importance of structural aspects of energy system models, spatial and temporal resolution, compared to uncertainties in input parameters such as final energy demand, discount rate and capital costs. We use global sensitivity analysis to uncover these interactions for two developing countries, Kenya, and Benin, which still lack universal access to electricity. We find that temporal resolution has a high influence on all assessed results parameters, and spatial resolution has a significant influence on the expansion of distribution lines to the unelectrified population. The larger overall influence of temporal resolution indicates that this should be prioritised compared to spatial resolution.

**Context & Scale**

Energy system models are used to understand the implications of different technology pathways and inform policy processes. This paper quantifies the importance of energy model design choices and the overall impact on the results for two developing countries, Kenya and Benin. The paper recommends that great care is needed when formulating the temporal resolution as this affects the overall insights of the modelling. We propose that the temporal and spatial resolution (when modelling network expansion), due to its influence on the results, should move towards being automated where also the structural parameters can be modified between scenarios, like other parametrical inputs such as cost parameters.

## GRAPHICAL ABSTRACT

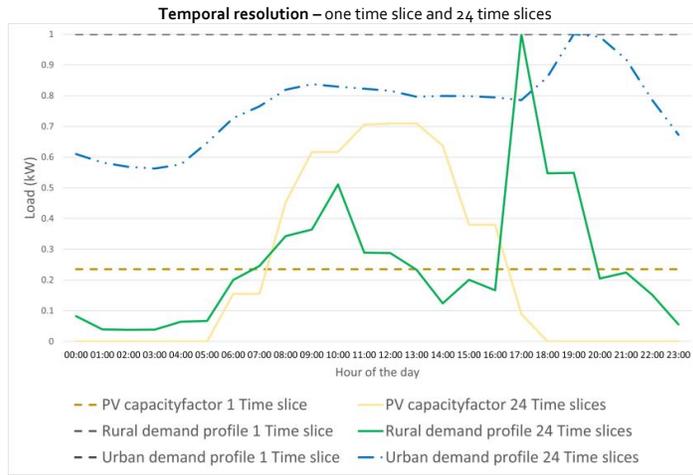

**Spatial resolution varied** 34 cells        67 cells                    100 cells

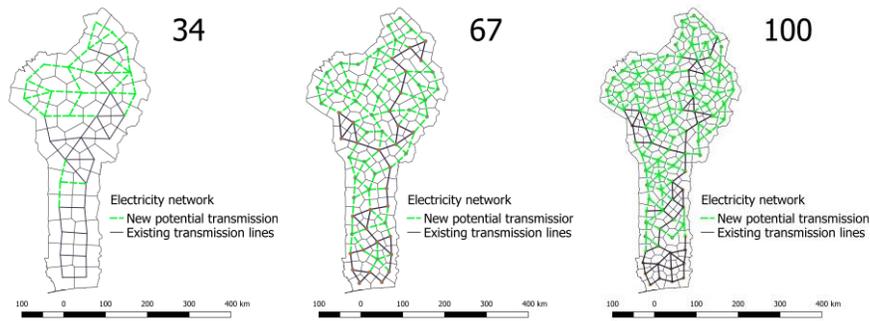

**Global sensitivity analysis results**

| Results parameter | Demand | Discount rate | Structural | | CF Error | Capital cost | | | | | | Fuel prices | | |
|---|---|---|---|---|---|---|---|---|---|---|---|---|---|---|
| Kenya Total Discounted Cost | 0.85 | 0.48 | 0.14 | 0.01 | 0.05 | 0.02 | 0.05 | 0.00 | 0.01 | 0.02 | 0.00 | 0.00 | 0.00 | 0.01 |
| Benin Total Discounted Cost | 0.89 | 0.39 | 0.07 | 0.01 | 0.04 | 0.02 | 0.08 | 0.00 | 0.00 | 0.01 | 0.00 | 0.00 | 0.03 | 0.00 |
| Kenya installed distribution lines | 0.88 | 0.04 | 0.18 | 0.16 | 0.12 | 0.10 | 0.22 | 0.04 | 0.00 | 0.00 | 0.00 | 0.00 | 0.01 | 0.01 |
| Benin installed distribution lines | 0.96 | 0.12 | 0.15 | 0.18 | 0.06 | 0.11 | 0.21 | 0.02 | 0.00 | 0.00 | 0.00 | 0.00 | 0.03 | 0.01 |
| Kenya Renewable electricity production share | 0.66 | 0.64 | 0.59 | 0.06 | 0.05 | 0.03 | 0.06 | 0.01 | 0.12 | 0.00 | 0.00 | 0.06 | 0.10 | 0.12 |
| Benin Renewable electricity production share | 0.70 | 0.33 | 0.36 | 0.11 | 0.28 | 0.28 | 0.29 | 0.06 | 0.05 | 0.02 | 0.06 | 0.07 | 0.20 | 0.04 |
| Input Parameters | Demand | Discount Rate | Temporal resolution | Spatial resolution | Capacity Factor error | Capital Cost PV | Capital Cost distribution lines | CapitalCost Battery kWh | Capital Cost Wind | Capital Cost distribution strengthening | CapitalCost transmission lines | Fuel price coal | Fuel price natural gas | Fuel price crude oil |

**INTRODUCTION**

In response to the evolving challenges facing the transition to a climate-neutral energy system, together with the need to provide universal access to modern energy services, energy system optimisation models (ESOMs) are moving towards larger, more complex and detailed models with a higher spatial and temporal resolution. This move is enabled by higher computational power and is driven by technological changes, such as the need to understand the emergence of intermittent renewable technologies, the expansion of transmission lines under different demands [1,2] and whether intermittent technologies, such as PV panels, play an important role in reaching universal access to modern energy by 2030 in countries with currently low electrification levels [3].

Despite the increasing detail, models are still a simplification of reality and require researchers to consider the uncertainties and errors introduced by simplifying assumptions [4]. Therefore, there is a need to consider the influence of these simplifying assumptions, both those that affect model structure, and the importance in relation to more commonly investigated model parameters, upon the results of the models. These can be expressed as structural and parametric uncertainties – and are therefore amenable to sensitivity analyses – to quantify the relative influence upon model results [5]. Sensitivity analyses can identify the importance of parameters in a model, while uncertainty analysis captures the most probable outcome based on the uncertainty and distribution range of the input parameters [6]. Global sensitivity analysis (GSA), as opposed to local sensitivity analysis, can capture interactions in linear and non-linear models [7]. ESOMs, even though they are often built upon linear programming (LP) [8], are not linear in their inputs and results [9].

There has been some progress in terms of understanding the model uncertainty of parameters in ESOMs and their importance in considering parametric uncertainty. Two models, Swiss-Energyscope [10,11] and EnergyScopeTD [12] are snapshot models of typical days and Calliope EU-28 [13] is a single-year ESOM. The GSA methods used in these studies are Morris [14] and Sobol' [7]. For multi-year ESOMs, we find UKTM [15], Balmorel [16,17] and ESME UK [18]. In these studies, the methods used were regression and Morris [14]. All of the aforementioned models are modelled for European high-income countries [19] such as the UK, Norway, and Belgium.

Many authors have investigated the influence on model results of changing either spatial or temporal resolution. The spatial resolution is varied to see the impact of spatial aggregation on the overall results [20–23]. The temporal resolution is aggregated using different methods such as typical day types [24,25] or clustering methods [24,26]. Preisman et al. [27] explored the accuracy and complexity of a power system optimisation model (PSOM) where the spatial and temporal resolution was varied with other parameters. Priesman et al. proposed reducing complexity in ESOMs first by grouping units (power plants and storage units), then temporal resolution and last spatial resolution. There has been less progress in understanding the relative importance of structural uncertainty for ESOMs. Yliruka et al. assessed in a GSA one structural parameter, spatial resolution, in a heat decarbonisation ESOM on a city level for the UK. The spatial resolution was compared to demand, efficiencies, technology cost and fuel prices, while the spatial resolution was found not to highly influence the total cost, it was identified as one of the most important factors for capacities of electricity, gas and heat networks [28].

Our review of the literature shows that with the increasing size of models in terms of spatiotemporal resolution, there is a need to better understand the influence on the results of structural assumptions and uncertainties relative to other model parameters. Furthermore, our literature search shows that there is a knowledge gap in developing countries of research in this field. Therefore, we ask the following research question: *What is the relative importance of the spatial and temporal resolution to demand, discount rate, cost of capital, capacity factor estimation error, and fuel costs in an ESOM for universal access to modern energy?*

We use two case applications, Kenya and Benin, to model universal access to electricity in the cost-minimising LP model generator called 'GEOSeMOSYS'[29]. Kenya is a country in eastern Africa and is endowed with large shares of renewable energy resources such as geothermal, solar, hydropower and wind and has ~29% of the population that still lacks access to electricity. Benin is a country in western Africa, which has almost 58% of the population who still lacks access to electricity. The power production runs mainly on fossil fuels and imports. For each country, we conduct a GSA in which 14 parameters vary, where the two structural parameters, spatial and temporal resolution, are assessed alongside the input parameters (see EXPERIMENTAL PROCEDURES for details on the modelled parameters and model setup). The spatial resolution is varied between one node to the smallest cell of ~1230 km² cells for Benin and ~5800 km² cells for Kenya and includes location-specific off-grid options as well as transmission expansion from the central power plants as seen in Figure 1 for Benin.

**34 cells**  **67 cells**  **100 cells**

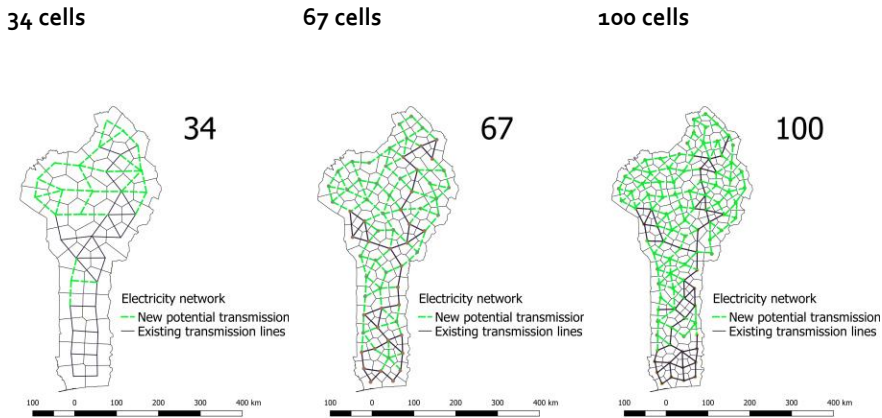

*Figure 1. The variation of the spatial resolution for Benin with the existing transmission lines (black lines) and evaluated expansion (green lines).*

The temporal resolution is resolved over one typical day and varies between 1 and 24 time slices (meaning hourly resolution for the day type at the highest resolution) with no seasonal variation (see EXPERIMENTAL PROCEDURES for more details on the spatial and temporal variations).

**RESULTS**

The objective function of the ESOM used in this study, Open Source Energy MOdel SYStem (OSeMOSYS), is to minimise the total discounted cost. The two most influential parameters for total discounted cost, meaning a higher score in $\mu^*_{i,SEE}$, are demand and discount rate followed by the capital cost of distribution lines (ranked fourth in Kenya and third for Benin) and temporal resolution (ranked fourth in Benin and third for Kenya) as seen in Figure 2.

| Results parameter | Demand | Discount rate | Structural | | | CF Error | Capital cost | | | | | | Fuel prices | | |
|---|---|---|---|---|---|---|---|---|---|---|---|---|---|---|---|
| Kenya Total Discounted Cost | 0.85 | 0.48 | 0.14 | 0.01 | | 0.05 | 0.02 | 0.05 | 0.00 | 0.01 | 0.02 | 0.00 | 0.00 | 0.00 | 0.01 |
| Benin Total Discounted Cost | 0.89 | 0.39 | 0.07 | 0.01 | | 0.04 | 0.02 | 0.08 | 0.00 | 0.00 | 0.01 | 0.00 | 0.00 | 0.03 | 0.00 |
| Kenya installed distribution lines | 0.88 | 0.04 | 0.18 | 0.16 | | 0.12 | 0.10 | 0.22 | 0.04 | 0.00 | 0.00 | 0.00 | 0.00 | 0.01 | 0.01 |
| Benin installed distribution lines | 0.96 | 0.12 | 0.15 | 0.18 | | 0.06 | 0.11 | 0.21 | 0.02 | 0.00 | 0.00 | 0.00 | 0.00 | 0.03 | 0.01 |
| Kenya Renewable electricity production share | 0.66 | 0.64 | 0.59 | 0.06 | | 0.05 | 0.03 | 0.06 | 0.01 | 0.12 | 0.00 | 0.00 | 0.06 | 0.10 | 0.12 |
| Benin Renewable electricity production share | 0.70 | 0.33 | 0.36 | 0.11 | | 0.28 | 0.28 | 0.29 | 0.06 | 0.05 | 0.02 | 0.06 | 0.07 | 0.20 | 0.04 |
| Input Parameters | Demand | Discount Rate | Temporal resolution | Spatial resolution | | Capacity Factor error | Capital Cost PV | Capital Cost distribution lines | CapitalCost Battery kWh | Capital Cost Wind | Capital Cost distribution strengthening | CapitalCost transmission lines | Fuel price coal | Fuel price natural gas | Fuel price crude oil |

*Figure 2. The influence of the 14 input parameters for Kenya and Benin is measured using the $\mu^*_{i,SEE}$ for three key result indicators: total discounted electricity system cost, number of km of installed distribution lines to unelectrified population and renewable electricity production share.*

For Benin, the unscaled statistical output can report on the actual effect ($\mu^*_i$), which amounts to 41 MUSD for the spatial resolution and the impact of changing the spatial resolution ranges between 1-3 % of the total discounted cost. The $\mu^*_i$ for spatial resolution for Kenya amounts to 1% of the mean total discounted cost. Comparing this to the literature, ranges of 1-2.75%[27] and ~9 %[21] and up to 23%[23] for the total discounted cost are observed. Focusing on the temporal resolution, the $\mu^*_i$ amounts to ~9% (254 MUSD) for Benin and $\mu^*_i$ amounts to ~14% (1,521 MUSD) for Kenya of the mean total discounted cost for each country.

Looking at the number of km installed distribution lines to the unelectrified population in the base year, the demand and cost of extending the distribution lines are still the most influential parameters ($\mu^*_{i,SEE}$), as the higher the demand the better the economies of scale are to meet the demand by expanding the grid. The spatial resolution is the third most important parameter ($\mu^*_{i,SEE}$) for Benin and the fourth most important for Kenya, and the higher spatial resolution leads to fewer installed distribution lines. This is partly related to the more realistic cost of extending the transmission lines to the adjacent cells which also leads to higher costs to connect them. This cost is not captured in the more generic one-node analysis where the transmission cost is applied per installed capacity.

Looking at the interaction/non-linearity effect (see EXPERIMENTAL PROCEDURES) between the parameters there is a very high interaction/non-linearity between all parameters except for demand for the expansion of the distribution lines. For total discounted cost the interaction/non-linearity is also high except for demand, capacity factor error, capital cost distribution strengthening, and discount rate. If this indicates an interaction, it means that there are second- or higher-order effects, so the change behind the result is explained by more than the primary input parameter (listed in Figure 2). A non-linear effect means that a change in the input parameter does not result in a linear change in model output.

The renewable electricity production (REP) share, for Kenya, is relatively high for all scenarios, ranging between 69%-90%, due to the large geothermal and hydropower production in the country. As seen in Figure 1, the most influential parameters for REP are the discount rate, demand, and temporal resolution, while for Benin the temporal resolution ranks second most influential and the discount rate third. The discount rate affects the REP share negatively when the discount rate is high, as the capital costs in the system are generally lower for fossil-fuelled supply options. These are preferred when the discount rate is high due to lower capital intensity compared to renewable technologies. For Benin, when the discount rate is low, hydropower is more cost-competitive than fossil generation.

The temporal resolution of 1 time slice introduces the propagation of errors in two ways. First, it is modelled as a base load supply as illustrated in Figure 3B where all demand *can* be supplied by a PV panel (regardless of nighttime or daytime) as compared to the 16 time slice scenario which is mainly supplied by PV with batteries in Kenya (Figure 3C). A similar trend is also observed for Benin. Second, the temporal resolution of one time slice omits the peak, leading the demand to be the same as the flat value (as illustrated in Figure 3A with the brown dashed line (average PV capacity factor) and black dashed line (demand profile)). When the peak is omitted, the total cost of the system is reduced as in OSeMOSYS the controlling factor for cost of capital is the modelled peak demand.

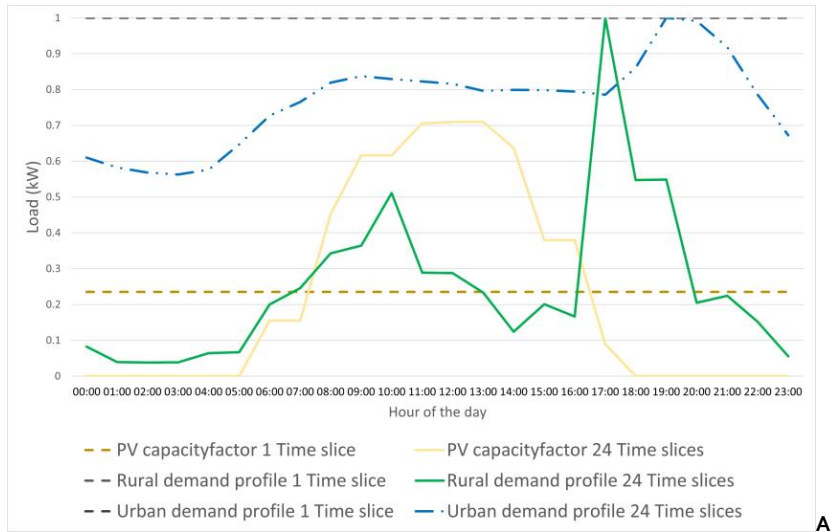

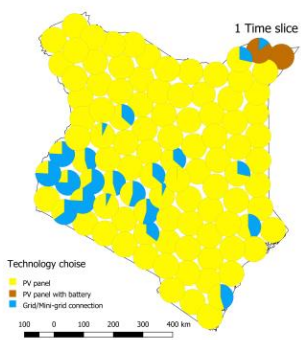

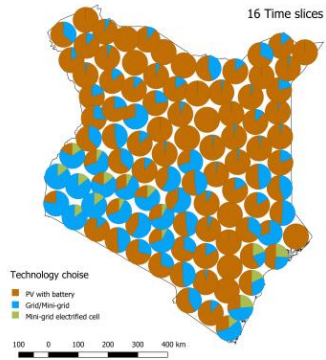

*Figure 3A-C. The figure illustrates the temporal resolution in the study from two perspectives. Figure A illustrates the daily demand profile with the blue dashed line representing the modelled urban demand, and the green line one of the rural demand profiles (Tier 4) in a 24-hour resolution. The grey dashed line represents the demand profile for one time slice. These demand profiles are compared to the PV supply profile in yellow (24-hour resolution) and dashed orange (one time slice). Figure B. illustrates for Kenya the results for one time slice analysis where the majority of the unelectrified population is supplied by a PV panel. Figure C. Illustrates for Kenya the results for one time slice analysis where the majority of the unelectrified population is supplied by a PV panel with a battery.*

**DISCUSSION**

We find that when varying the structural parameters, temporal and spatial resolution, the temporal resolution is more influential compared to the spatial resolution in this study. Spatial resolution has a high influence on the distribution line's expansion, but a more moderate impact on the total cost and renewable electricity production. The higher importance for temporal resolution could be explained by two factors: 1) the averaged temporal resolution leads to the one time slice resolution having a much lower peak than the higher resolution. The peak is the controlling variable for the capacity requirements, upon which the capital cost of the system is based. 2) The most preferred technology to the unelectrified population, when demand levels are low, is a PV panel with or without battery, which has a quite low spatial variation. A homogenous variation of energy potential over the studied area can motivate a lower spatially resolved model[30]. At the same time a lower resolved model does not capture green field expansion of the transmission lines, omitting more exact transmission and distribution costs.

The spatial resolution does influence the expansion of distribution lines for both countries, implying that it cannot be 'fixed'. Furthermore, due to the linearity assumption, very small amounts of transmission lines can be installed at quite a low cost. If the transmission grid would be implemented as a mixed integer linear programming, allowing only to be installed at a certain capacity, then these 'barriers' could potentially influence the results to a higher impact, however, this needs further investigation.

The method of Morris is a computationally efficient method to understand sensitivities in a model. However, considering the interactions/non-linearity observed for many of the input parameters indicates that further investigation is needed. Furthermore, the scaled measure of sensitivity $\mu^*_{i,SEE}$ for the distribution line length shows shifts in relative importance when compared to the $\mu^*_i$ for spatial resolution. This highlights the limitation of the method of Morris, which can only give indications of influences, especially to non-influential parameters. To further uncover second- and higher-order interactions, other sensitivity measures such as Soboł can be used. This, however, comes at a large computational expense as Soboł number of runs is N($k$+2), where N≈500-1000 and $k$ are the number of input parameters[7], leading to at least 8000 scenario runs for this study. For context, running the two countries, Benin and Kenya, for the 150 scenarios per country for this study took 66 hours on a 250 GB RAM computer.

The demand is highly influential, and the levels of Tier 1 to Tier 4 for the unelectrified population in the base year is a large range. Tier 4 is equivalent to a household with high-power appliances, while Tier 1 is a household with some lighting and cell phone charging. The range can therefore be discussed if it is likely, however, with an ambitious government it could be plausible. The demand profile for the electrified demand was assumed to have the same profile for both countries, and it remained the same throughout the modelling. This is a simplification, and better data to enrich the electrified demand would give better insight into these interactions as well. As the focus of this study was to understand the pathways to universal access to electricity, the demand profiles for the unelectrified population in the base year were the focus and varied with the corresponding demand level.

The data quality also affects the results, and the transmission and distribution network dataset for Kenya is more detailed compared to Benin (see Supplementary material), therefore, the open-access data used in this analysis can also influence the results when calculating the need for new distribution lines. The capacity factor error also affected the results; however, the error was implemented for the full year either up or down. It was applied to wind power as well as solar PV, with an error of 5%. A case study in Norway[31] found that the wind farm estimation from Renewables Ninja[32] could have a much higher error margin.

**Conclusions and future work**

This paper highlights the influence of the structural parameters, spatial and temporal, in an ESOM for electricity access in Kenya and Benin. We find that the two structural parameters influence the results, however, the temporal resolution has a higher impact than the spatial resolution in both case applications. This finding is important since both the spatial and temporal resolution are often set early in the modelling process and the subsequent processing of data depends on these. They are often not easy to change and therefore have a lock-in effect. We propose that especially when modelling the expansion of the network, spatial and temporal resolution should be customisable and changed, even between scenarios, depending on the research question on hand.

It was also observed that, compared to other studies[21,23,27], the importance of the spatial resolution varies and therefore the technology mix (e.g., high penetration of PV), and other parameters in the model (e.g., congestion in the transmission lines, limits to expansion of the network) could be explanations of the variance of the importance of the spatial resolution between the different studies and this paper.

Proper representation of the demand peak is important, and therefore implementations to better represent it can help mitigate the influence of temporal resolution if modelled in a lower temporal resolution, e.g. through temporal clustering or heuristic methods[24,34], or by adding a peak parameter to the ESOM to secure the overall peak of the system (leading to higher capital cost investments).

In this GSA, there was no seasonal representation which can influence the results where there are seasonal changes e.g., hydrology, wind power, demand. Furthermore, the spatial resolution in this study was based on making the area as even as possible. Using other clustering approaches can help to simplify without affecting the results as seen in [20]. Therefore, including spatiotemporal clustering/aggregation approaches in the GSA would prove valuable to understanding that the implementation interacts in the intended way. Apart from the structural parameters, we can also see that demand and discount rates are highly influential, and a more detailed representation of these can better help us understand the evolution of the system. This can be done by first quantifying the uncertainty surrounding the most influential variables and then exploring the consequences of this uncertainty range. Agutu et al.[35] suggest that off-grid solutions have a much higher cost of capital than other technology options. In this study we use a uniform discount rate, however, a technology-specific discount rate could alter the results, and thus would also be a suitable future parameter to include.

**EXPERIMENTAL PROCEDURES**

**Resource Availability**

*Lead Contact*
Further information and requests for resources and materials should be directed to and will be fulfilled by the lead contact, Nandi Moksnes (Nandi@moksnes.se).

*Materials Availability*
No materials were used in this study.

*Data and Code Availability*
The source code used in this study is available at GSA_Spatial_temporal: https://github.com/KTH-dESA/GSA_Spatial_temporal.

**Country model data**

To address the paper's research question we use the open-source ESOM *OSeMOSYS* [36]. OSeMOSYS has recently been applied in a high spatial resolution model for electricity access in Kenya[29]. Moksnes et al. also provides an open-source model generator[37], *GEOSeMOSYS*, that the spatial resolution in OSeMOSYS to be varied programmatically. GEOSeMOSYS is spatially resolved in the sense that it consists of distributed electricity supply options as well as extending the existing grid to remote locations without electricity access in a multi-year analysis including the central power plant optimisation. For this paper, we develop the open-source code of GEOSeMOSYS further to also include varying temporal resolution, demand, discount rate, fuel, capacity factor error, and capital cost to run a GSA (see Table 1). To understand how the model is further spatially disaggregated we recommend the reader read the paper[29].

To obtain more generalisable results, two countries with low electrification rates are modelled: Kenya (29% unelectrified population) and Benin (58% unelectrified population) [38]. The general powerplant structure and future potentials, which is a pre-requisite to the GEOSeMOSYS model generator, are extracted from TEMBA[39] for the Benin electricity sector (excluding trade with neighbouring countries) and Kenya the model from the paper[29].

**Method of Morris**

The spatial and temporal resolution in ESOMs drives the computational time as the higher the resolution the larger the matrix of constraints. Therefore, the driving factor of the analysis is to understand at what level can the model be reduced without compromising the results of the model. In GSA three outputs can be generated using different GSA algorithms. *Factor prioritisation*, which describes the most influential parameters where most effort and research should be given, *factor fixing*, which identifies non-influential parameters that can be removed to simplify the model, and *factor mapping*, where certain regions of interest are identified and what input parameters that drive these outcomes [7].

The Method of Morris provides a basic factor prioritisation. It performs well when the number of uncertain factors is high and the computational time is long[40]. It is most well suited for factor fixing type of analysis. It belongs to the One-At-the-Time (OAT) type of analysis, where variations around the base point are assessed, however, it overcomes some of the shortcomings that OAT has (such as the inability to capture non-linear interactions) by changing the values of all other parameters. By applying averages to different local measures, the interactions between the input parameters can be captured[7].

The method follows *trajectories* (j), and each trajectory (10 in this study) is calculated (*k*+1) times, where *k* represents the number of input parameters. At each step of the m-th trajectory, the change in results (elementary effect) for each input parameter (i) to the j-th results parameter (Y) is calculated. To be able to compare the different results to each other (with different orders of magnitude), scaling the elementary effects as in Sin and Gearney[41] is applied. This means that the standard deviation of the input parameter is divided by the standard deviation of the results parameter per trajectory, to act as a scaling factor (Equation 1 and Equation 2).

*Equation 1. Scaled elementary effect per trajectory[10,41]*

$$SEE_{ij}^m = \frac{\delta Y_j}{\delta x_i}\frac{\sigma_{x_i}}{\sigma_{Y_j}}$$

*Equation 2. The absolute mean scaled elementary effect per input parameter[10,41]*

$$\mu_{i,SEE}^* = \frac{1}{r}\sum_{m=1}^{r}|SEE_i^m|$$

To further understand the non-linear/interactions between the parameters, the standard deviation of the elementary effect $\sigma_i^2$ and the absolute mean elementary effect $\mu_i^*$ are retrieved and divided with each other (Equation 3, Equation 4 and Equation 5).

*Equation 3. The absolute elementary effect per trajectory[40]*

$$\mu_i^* = \frac{1}{r} \sum_{m=1}^{r} \left|\frac{\delta Y_j}{\Delta}\right|$$

Where Δ step is set to 2/3 in this study.

*Equation 4 Standard deviation of the elementary effect*

$$\sigma_i^2 = \frac{1}{r-1} \sum_{m=1}^{r} \left(\frac{\delta Y_j}{\Delta} - \mu\right)^2$$

*Equation 5 Interaction nonlinearity measure*

$$\frac{\sigma_i}{\mu_i^*} > 1, nonlinearity\ or\ interaction\ with\ other\ parameters$$

In this paper, we want to understand if an ESOM can be reduced in size (factor fixing) without affecting the overall results. We, therefore, use the Method of Morris[14,40] using the open-source sensitivity Python library SALib[42]. To allow for calculations on the scaled elementary effects an addition was made to the SALib library[43].

The GSA includes the two structural parameters spatial and temporal resolution, as seen in Table 1, and 12 input parametrical uncertain parameters including demand, discount rate, various costs and errors in capacity factor estimates for solar and wind. The input parameters ranges, from which the samples are generated, are based on a literature review to find min and max values to find a plausible solution space. All parameters are assumed to be independent of each other in the sensitivity analysis, and therefore parameters that co-vary are only represented once in the input parameters list. The two demand parameters: demand profile and demand level, are assessed not to be independent of each other, and therefore they will take the same sample value for each trajectory. Similarly, are PV panels for residential and mini-grid, and heavy fuel oil and diesel modelled in the GSA as one uncertain parameter. The modelling period is 2020-2040 for all scenarios.

*Table 1 Sensitivity parameters in the Method of Morris workflow*

| Type | Input parameter | Value range | Source | Unit |
|---|---|---|---|---|
| Structural parameter | Spatial | One node → ~1230 km² for Benin ~5800 km² for Kenya | Limited by the size of the model | Number of nodes |
| | Daily temporal resolution | 1 → 24 | Limited by the granularity of the hourly resolution | Number of evenly spread average slices |
| Input parameter | Demand Unelectrified | Multi-tier framework Tier 1 → 4 | [44, 45] | GJ |
| | Discount Rate | Kenya: 8% → 21% Benin: 7% → 20% | [35] | Percent |
| | Transmission Line Cost | 2 → 4 | [46] | USD/kW-km |
| | Capital Cost distribution strengthening cost | 50 → 400 | Assumed value | USD/kW |
| | Distribution Line Cost | 10,000 → 28,000 | [47] | USD/km |
| | Capital Cost residential stand-alone PV panel | 2,743 → High:1,651/Low: 671 | [48] | USD/kW |
| | Capital Cost residential Battery (kWh) | 1339 (constant) + 685 → High: 1339 (constant) + 439/Low: 1339 (constant)+ 277 | [48] with adjusted constant to Kenyan wages [49] | USD/kWh |
| | Capital Cost Wind power | 1405 → High: 913/Low: 589 | [48] | USD/kW |
| | Capacity Factor PV panel/Wind power | +- 5 | [51] | percent |
| | Fuel price Natural Gas | 9 → High : 8, Low: 4 | [52] | USD/GJ |
| | Fuel price Diesel | 17 → High:14, Low: 7 | [52] | USD/GJ |
| | Fuel price Coal | 5 → High: 2.9 Low: 2.1 | [52] | USD/GJ |

**Selection of model outputs**

The results parameters chosen for this study were three: Total discounted cost, the number of installed km of distribution lines to the unelectrified population in the base year and the renewable electricity production. The total discounted cost is interesting as this is the objective function of the model and gives an understanding of what optimal solution would be preferred. The number of installed km to the unelectrified is chosen as it indicates the economic viability of connecting unelectrified households to grid/minigrid as opposed to rooftop PV panels.

Finally, as the share of renewable electricity production also is part of the SDG7, the parameters that are highly influential are important to capture to reach the goals.

**Spatial resolution - Spatial clusters**

The spatial cells are divided using QGIS to form even cells over the countries. No specific aggregation, apart from the spatial resolution is modelled. The first step is generating 10,000 - 100,000 random points in the administrative boundary. The random points are then clustered using k-mean[1], aggregated and then the centroid is extracted from these polygons. Finally, Voronoi cells are generated from the centroid and then clipped by the administrative boundaries. For Kenya, the spatial resolution is ~5800 km² and for Benin ~1230 km².

**Demand levels**

For the demand of the electrified population, the estimated low and Vision scenario is used for Kenya[29] and for Benin the two demand trajectories[39] are applied. The unelectrified population in the base year follow the Multi-Tier framework of Tier 1- to Tier 4[44] and the corresponding demand profiles [45]. The multi-tier framework gives a more detailed perspective on the electrification of unelectrified households. Tier 1 is like having an electric lantern in the house and Tier 4 corresponds to a fully electrified household with 75% reliance on continuous supply.

**Temporal resolution**

The temporal resolution is divided into even daily average splits, with no specific adjustment to the intermittent supply options or demand profiles, and only one season. The temporal resolution varies from 1 to 24 timesteps per day (i.e., hourly at the highest granularity), and as illustrated in Figure 4, one can see that the average value of one time slice is flat throughout the day while the 24 time slices follow the original data set most closely.

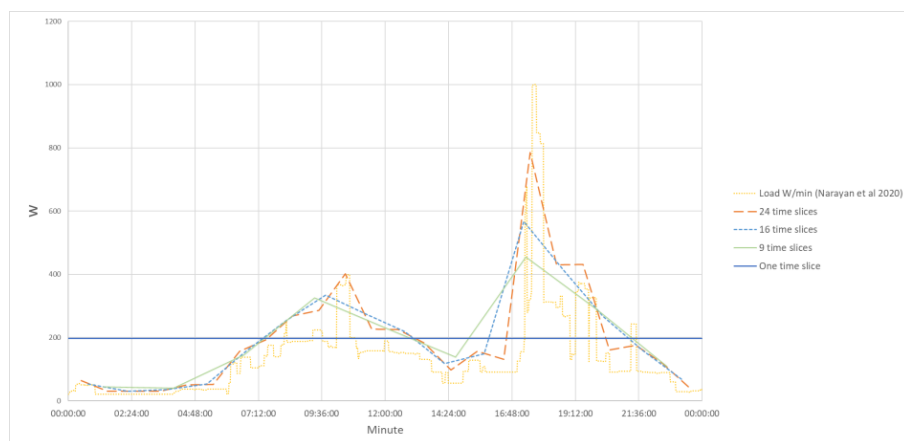

*Figure 4. Example of time-slice representation, here for demand profile for Tier 4 level demand compared to the source dataset from Narayan et al.[45]*

**AUTHOR CONTRIBUTIONS (CRediT)**

Conceptualization: N.M, W.U. Methodology: N.M, W.U. Software: N.M Validation: N.M, W.U., Formal Analysis: N.M., Investigation: N.M., W.U., Data curation: N.M. Writing – original draft: N.M. Writing – Review & Editing: N.M., W.U., Supervision: W.U. Project Administration: N.M.

**DECLARATION OF INTERESTS**

---

[1] In QGIS version 3.18.

The authors declare no competing interests.

**Supplementary material**

**Transformer and household connection data**

The transformer's cost and household connection calculations were an addition to this version of the model generator: GEOSeMOSYS. It was based on the calculation from van Ruijven et al. [53] in a 1x1 km resolution for both Benin and Kenya. The estimated number of LV networks per MV line is estimated through the minimum of either the number of households or the maximum of the estimated LV line length or estimated LV capacity. In addition, one MV-line is expected to be connected to the 1x1km settlement, and a connection is added to each one. The sum of all estimated transformers for Kenya and Benin was then multiplied by a transformer cost of 3500 USD[54]. In addition, a connection cost per household was added of 125 USD/connection[55]. To not create a lump sum per larger cell, the estimated cost was divided by the estimated km of the distribution line[29] and added to the distribution line cost which is multiplied by the number of km installed lines. For Kenya, the average cost was 5646 USD/km and for Benin 7448 USD/km.

**Electricity network data**

The electricity network that is used in the study is illustrated in the below figure.

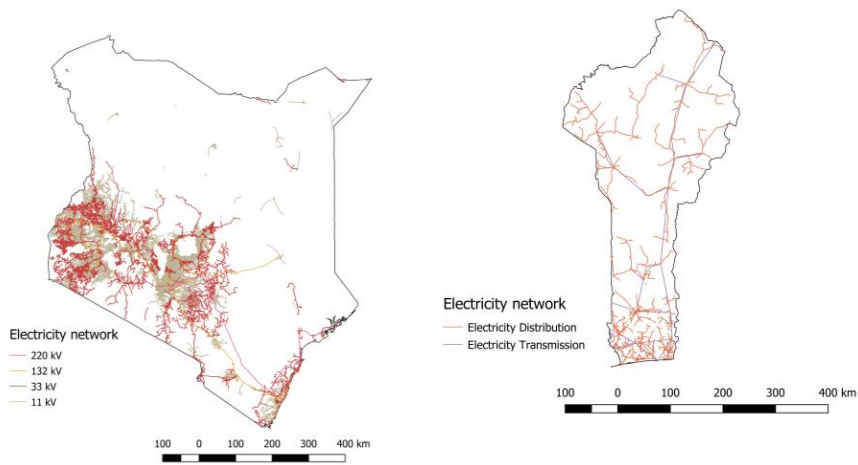

*Figure. GIS data set for the electricity network of Kenya[56] (left) and Benin[57,58] (right).*